\documentclass[runningheads]{llncs}
\usepackage{graphicx}
\usepackage{listings}
\usepackage{xcolor}
\usepackage{hyperref}

\hypersetup{
    colorlinks=true,
    linkcolor=blue,
    filecolor=magenta,      
    urlcolor=cyan,
}
\lstdefinestyle{Python}{
    language        = Python,
    basicstyle      = \ttfamily,
    keywordstyle    = \color{blue},
    keywordstyle    = [2] \color{teal}, % just to check that it works
    stringstyle     = \color{green},
    commentstyle    = \color{red}\ttfamily
}

\begin{document}
\lstset{
    frame       = single,
    numbers     = left,
    showspaces  = false,
    showstringspaces    = false,
    captionpos  = t,
    caption     = \lstname
}
\title{Introduction To Medical Image Registration with DeepReg, Between Old and New}
\titlerunning{Introduction to Medical Image Registration with DeepReg}
% If the paper title is too long for the running head, you can set
% an abbreviated paper title here
%
\author{Nina Montana Brown\inst{1, 2}\orcidID{0000-0001-5685-971X} \and
Yunguan Fu\inst{1, 2, 3}\and
Shaheer Saeed\inst{1, 2}\and
Adria Casamitjana \inst{2} \and
Zachary M. C. Baum \inst{1, 2}\and
Remi Delaunay \inst{1, 4} \and
Qianye Yang \inst{1,2}\and
Alexander Grimwood \inst{1,2} \and
Zhe Min \inst{1} \and
Ester Bonmati \inst{1,2} \and
Tom Vercauteren \inst{4}\and
Matthew J. Clarkson \inst{1,2} \and
Yipeng Hu\inst{1,2}}
\authorrunning{N. Montana Brown et al.}
% First names are abbreviated in the running head.
% If there are more than two authors, 'et al.' is used.
%
\institute{Wellcome/EPSRC Centre for Surgical and Interventional Sciences, University College London \and
Centre for Medical Image Computing, University College London
 \and
InstaDeep \and Department of Surgical and Interventional Engineering, King’s College London}
\maketitle              % typeset the header of the contribution
\begin{abstract}
This document outlines a tutorial to get started with medical image registration using the open-source package DeepReg. The basic concepts of medical image registration are discussed, linking “classical" methods to newer methods using deep learning. Two iterative, classical algorithms using optimisation and one learning-based algorithm using deep learning are coded step-by-step using DeepReg utilities, all with real, open-accessible, medical data.

\end{abstract}
\section{Objective of the Tutorial}
This tutorial introduces a new open-source project \href{https://github.com/DeepRegNet/DeepReg}{DeepReg}, currently based on the latest release of TensorFlow 2. This package is designed to accelerate research in image registration using parallel computing and deep learning by providing simple, tested entry points to pre-designed networks for users to get a head start with. Additionally, DeepReg provides more basic functionalities such as custom TensorFlow layers which allow more seasoned researchers to build more complex functionalities. \par
A previous MICCAI workshop learn2reg provided an excellent example of novel algorithms and interesting approaches in this active research area, whilst this tutorial explores the strength of the simple, yet generalisable design of DeepReg.
\begin{itemize}
    \item Explain basic concepts in medical image registration
    \item Explore the links between the modern algorithms using neural networks and the classical iterative algorithms (also using DeepReg);
    \item Introduce the versatile capability of DeepReg, with diverse examples in real clinical challenges.
\end{itemize}

Since DeepReg has a pre-packaged command line interface, minimum scripting and coding experience with DeepReg is necessary to follow along this tutorial. Accompanying the tutorial is the DeepReg documentation and a growing number of demos using real, open-accessible clinical data. This tutorial will get you started with DeepReg by illustrating a number of examples with step-by-step instructions.

\subsection{Set-up}
This tutorial depends on the package \href{https://github.com/DeepRegNet/DeepReg}{DeepReg}, which in turn has external dependencies which are managed by pip. The current version is implemented as a TensorFlow 2 and Python$\geq$3.7 package. We provide this tutorial in full as a \href{https://colab.research.google.com/github/DeepRegNet/DeepReg/blob/main/docs/Intro_to_Medical_Image_Registration.ipynb}{Jupyter notebook} with the source code, submitted as a \href{http://www.miccai.org/education/educational-challenge/}{MICCAI Educational Challenge}.
\par
Training DNNs is computationally expensive. We have tested this demo with GPUs provided by Google through Google Colab. Training times have been roughly measured and indicated where appropriate. You can run this on CPU but we have not tested how long it would take. \par

Firstly, we make some folders to store the outputs of the tutorial.

\begin{lstlisting}[style = Python ,numbers=none, caption = {Making initial folders to store tutorial code}]
import os

os.chdir("/content")
# Make a directory "MICCAI_2020_reg_tutorial"
if not os.path.exists("./MICCAI_2020_reg_tutorial"):
  os.makedirs("./MICCAI_2020_reg_tutorial")
# Move into the dir
os.chdir("./MICCAI_2020_reg_tutorial")
print(os.getcwd())

\end{lstlisting}
Now we set up these dependencies by installing DeepReg. This may take a few minutes. You may need to restart the runtime first time installing and there might be a few version conflicts between the pre-installed datascience and deepreg libraries - but these are not required in this tutorial.

\begin{lstlisting}[style = Python,numbers=none, caption={Installing DeepReg and it's dependencies}]
# Clone the DeepReg repository which contains the code
! git clone https://github.com/DeepRegNet/DeepReg
%cd ./DeepReg/
# Switch to a fixed version
! git checkout tags/miccai2020-challenge
# pip install into the notebook env
! pip install -e .
print(os.getcwd())
import yaml
yaml.__version__  # if necessary, first update the yaml
# version such that the config files can be 
# correctly parsed
#(this may require restart the runtime)
! pip install -U PyYAML
 
# We import some utility modules.
import nibabel
import tensorflow as tf 
import deepreg.model.layer as layer
import deepreg.model.loss.image as image_loss
import deepreg.model.loss.deform as deform_loss
import deepreg.model.layer_util as layer_util
import matplotlib.pyplot as plt
import os
import h5py
import numpy as np
from tensorflow.keras.utils import get_file

# We set the plot size to some parameters.
plt.rcParams["figure.figsize"] = (100,100)
print(os.getcwd())
str = "/content/MICCAI_2020_reg_tutorial/DeepReg"
if not os.getcwd() == str:
  os.chdir(str)
  print(os.getcwd())

\end{lstlisting}

\section{Introduction to Registration}
Image registration is the mapping of one image coordinate system to another and can be subdivided into rigid registrations and non-rigid registrations, depending on whether or not higher-dimensional tissue deformations are modeled as opposed to, for example, a 6 degree-of-freedom (3 translational axes + 3 rotational axes) rigid transformation. Data may be aligned in many ways - spatially or temporally being two key ones. Image registration is an essential process in many clinical applications and computer-assisted interventions \cite{Haskins20}\cite{Hill01}. \par

Applications of medical image registration include - but are not limited to:\par
\begin{itemize}
    \item Multi-modal registration for image-guided surgery: for example, aligning real-time ultrasound scans to pre-operative CT or MRI scans to real-time achieve guidance in abdominal applications \cite{Hu12}\cite{Ramalhinho18}.
    \item Atlas-based image segmentation: aligning new images to those carefully segmented, such that the reference segmentations can be propagated to those new images \cite{Valdes02}.
    \item Longitudinal comparison of images for a given patient with the same imaging modality: for example, comparing the outcome of given cancer treatment in a patients' scans over time \cite{Cazoulat16}\cite{Yang20}.
    \item Inter-subject comparison: for example, a population study of organ shapes \cite{Hu15}.
\end{itemize}

Typically, we refer to one of the images in the pair as the moving image and the other as the fixed image. The goal is to find the correspondence that aligns the moving image to the fixed image - the transform will project the moving coordinates into the fixed coordinate space. The correspondence specifies the mapping between all voxels from one image to those from another image. The correspondence can be represented by a dense displacement field (DDF) \cite{Ashburner07}, defined as a set of displacement vectors for all pixels or voxels from one image to another. By using these displacement vectors, one image can be "warped" to become more "similar" to another. \par
\begin{figure}
    \centering
    \includegraphics[width=\textwidth]{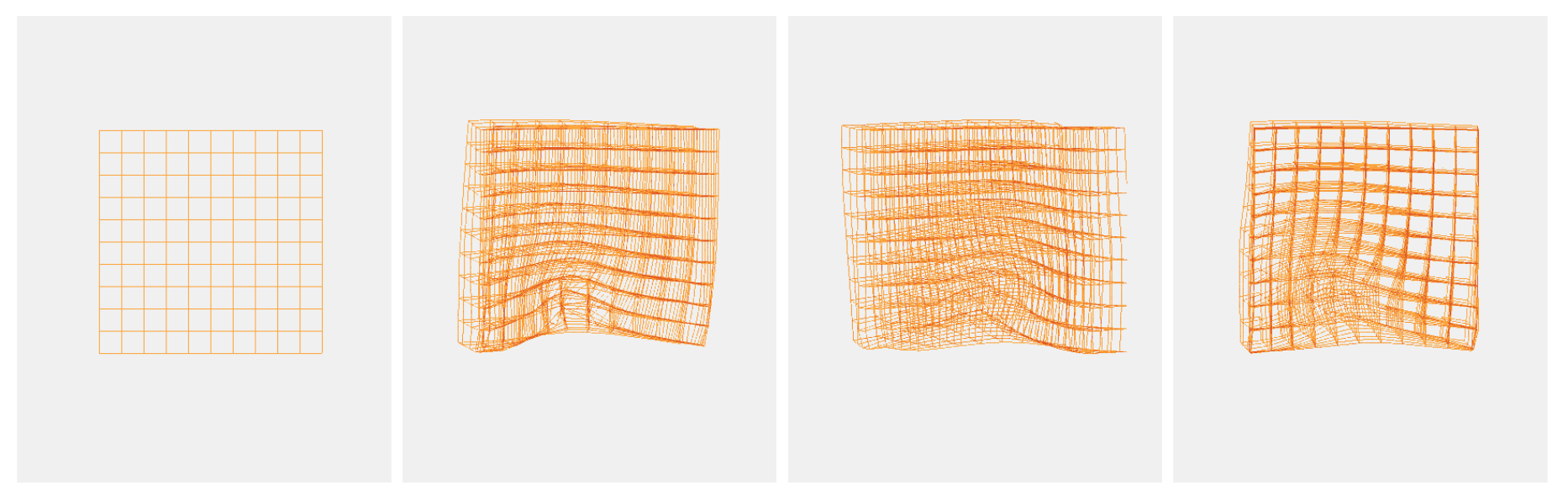}
    \caption{A 3D grid, representing voxels in a 3D image volume, is being warped by displacing each voxel position.}
    \label{fig:displacement}
\end{figure}

Fig. ~\ref{fig:displacement} shows a non-rigid example.\par

\subsection{Classical vs Learning Methods}
Image registration has been an active area of research for decades. Historically, image registration algorithms posed registration as an optimization problem between a given pair of moving and fixed images. In this tutorial, we refer these algorithms as to the classical methods - if they only use a pair of images, as opposed to the learning-based algorithms, which require a separate training step with many more pairs of training images (just like all other machine learning problems).

\subsubsection{Classical Methods}
In the classical methods, a pre-defined transformation model, rigid or nonrigid, is iteratively optimised to minimize a similarity measure - a metric that quantifies how "similar" the warped moving image and the fixed image are. \par
Similarity measures can be designed to consider only important image features (extracted from a pre-processing step) or directly sample all intensity values from both images. As such, we can subdivide algorithms into two sub-types:
\begin{itemize}
    \item \textbf{Feature-based registration:} Important features in images are used to calculate transformations between the dataset pairs. For example, point set registration - a type of features widely used in many applications - finds a transformation between point clouds. These types of transformations can be estimated using Iterative Closest Point (ICP) \cite{Besl92} or coherent point drift \cite{Myronenko10} (CPD), for rigid or nonrigid transformation, respectively. \par
    For example, the basis of ICP is to iteratively minimise the distance between the two point clouds by matching the points from one set to the closest points in the other set. The transformation can then be estimated from the found set of corresponding point pairs and repeating the process many times to update the correspondence and the transformation in an alternate fashion.
    \item \textbf{Intensity-based registration:} Typically, medical imaging data does not come in point cloud format, but rather, 2D, 3D, and 4D matrices with a range of intensity values at each pixel or voxel. As such, different measures can be used directly on the intensity distributions of the data to measure the similarity between the moving and fixed images. Examples of measures are cross-correlation, mutual information, and simple sum-square-difference - these intensity-based algorithms can optimize a transformation model directly using images without the feature extraction step.

\end{itemize}

Many of today's deep-learning-based methods have been heavily influenced - and derived their methods from - these prior areas of research.

\subsubsection{Why use Deep Learning for Medical Image Registration?}
Usually, it is challenging for classical methods to handle real-time registration of large feature sets or high dimensional image volumes owing to their computationally intense nature, especially in the case of 3D or high dimensional nonrigid registration. State-of-the-art classical methods that are implemented on GPU still struggle for real-time performance for many time-critical clinical applications. \par
Secondly, classical algorithms are inherently pairwise approaches that can not directly take into account population data statistics and relying on well-designed transformation models and valid similarity being available and robust, challenging for many real-world tasks. \par
In contrast, the computationally efficient inference and the ability to model complex, non-linear transformations of learning-based methods have motivated the development of neural networks that infer the optimal transformation from unseen data \cite{Haskins20}. \par
However, it is important to point out that:
\begin{itemize}
    \item Many deep-learning-based methods are still subject to the limitations discussed with classical methods, especially those that borrow transformation models and similarity measures directly from the classical algorithms;
    \item Deep learning models are limited at inference time by how the model was trained - it is well known that deep learning models can overfit to the training data;
    \item Deep learning models can be more computationally intensive to train than classical methods at inference;
    \item Classical algorithms have been refined for many clinical applications and still work well.
\end{itemize}

\section{Registration with Deep Learning}
In recent years, learning-based image registration has been reformulated as a machine learning problem, in which, many pairs of moving and fixed images are passed to a machine learning model (usually a neural network nowadays) to predict a transformation between a new pair of images. \par
In this tutorial, we investigate three factors that determine a deep learning approach for image registration:
\begin{itemize}
    \item What type of network output is one trying to predict?
    \item What type of image data is being registered? Are there any other data, such as segmentations, to support the registration?
    \item Are the data paired? Are they labeled?
\end{itemize}

\subsection{Types of Network Output}
We need to choose what type of network output we want to predict.
\par
\subsubsection{Predicting a dense displacement field}
Given a pair of moving and fixed images, a registration network can be trained to output dense displacement field (DDF) \cite{Ashburner07} of the same shape as the moving image. Each value in the DDF can be considered as the placement of the corresponding pixel / voxel of the moving image. Therefore, the DDF defines a mapping from the moving image's coordinates to the fixed image.
In this tutorial, we mainly focus on DDF-based methods.
\subsubsection{Predict a static velocity field}

Another option is to predict a static dense velocity field (SVF or DVF) between a pair of images, such that a diffeomorphic DDF can be numerically integrated. We refer you to \cite{Ashburner07} and \cite{Vercauteren09} for more details.
\subsubsection{Predict an affine transformation}

A more constrained option is to predict an affine transformation and parameterize the affine transformation matrix to 12 degrees of freedom. The DDF can then be computed to resample the moving images in fixed image space.
\subsubsection{Predict a region of interest}

Instead of outputting the transformation between coordinates, given moving image, fixed image, and a region of interest (ROI) in the moving image, the network can predict the ROI in fixed image directly. Interested readers are referred to the MICCAI 2019 paper \cite{Hu19}.

\subsection{Data Availability, level of supervision, and network training strategies}
Depending on the availability of the data labels, registration networks can be trained with different approaches. These will influence our loss choice.
\subsubsection{Unsupervised}
When multiple labels are available for each image, the labels can be sampled during training, such that only one label per image is used in each iteration of the data set (epoch). We expand on this for different dataset loaders in the DeepReg dataset loader API but do not need this for the tutorials in this notebook. \par

The loss function often consists of the intensity-based loss and deformation loss.
\begin{figure}
    \centering
    \includegraphics[width=\textwidth]{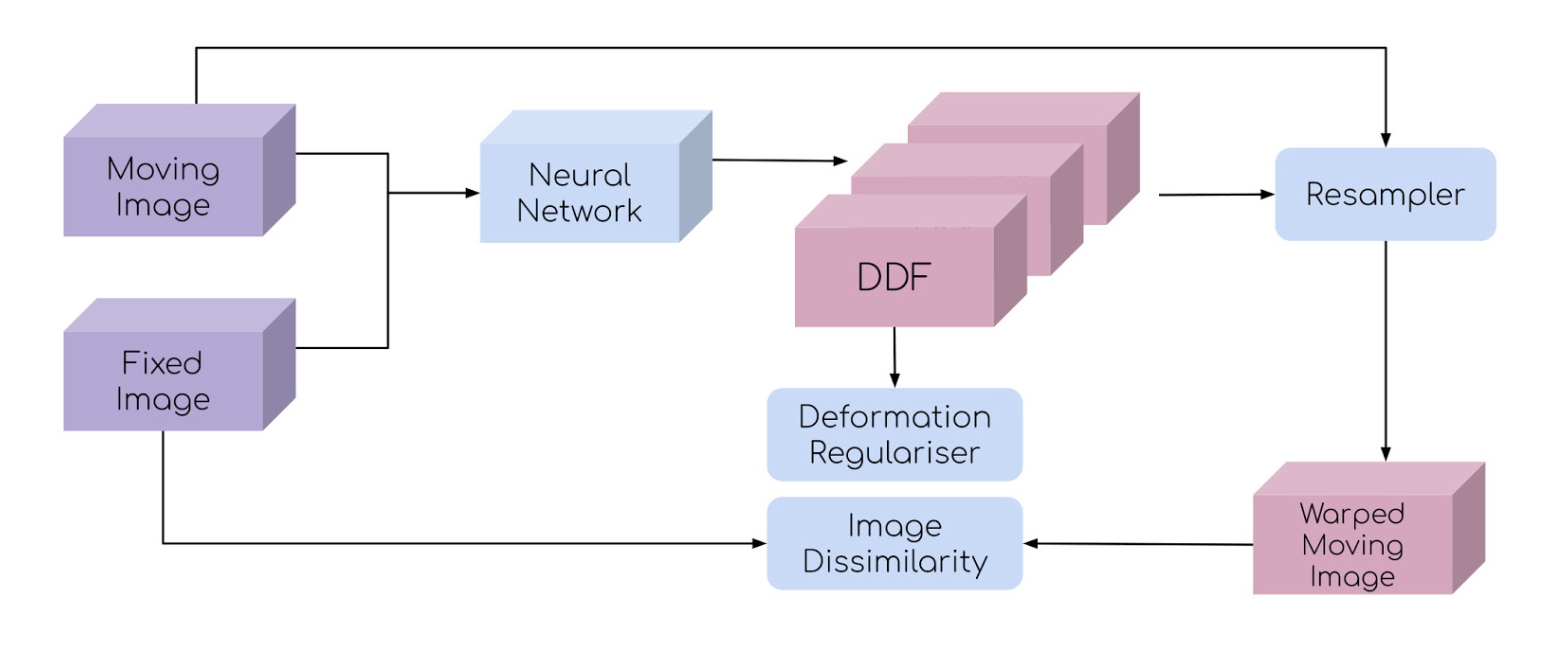}
    \caption{A schematic illustration of an unsupervised image registration network}
    \label{fig:unsup}
\end{figure}

\subsubsection{Weakly-supervised}
When an intensity-based loss is not appropriate for the image pair one would like to register, the training can take a pair of corresponding moving and fixed labels (in addition to the image pair), represented by binary masks, to compute a label dissimilarity (label based loss) to drive the registration.\par
Combined with the regularisation on the predicted displacement field, this forms a weakly-supervised training. An illustration of a weakly-supervised DDF-based registration network is provided below.\par
When multiple labels are available for each image, the labels can be sampled during training, such that only one label per image is used in each iteration of the data set (epoch). Details are again provided in the DeepReg dataset loader API but not required for the tutorials.

\begin{figure}
    \centering
    \includegraphics[width=\textwidth]{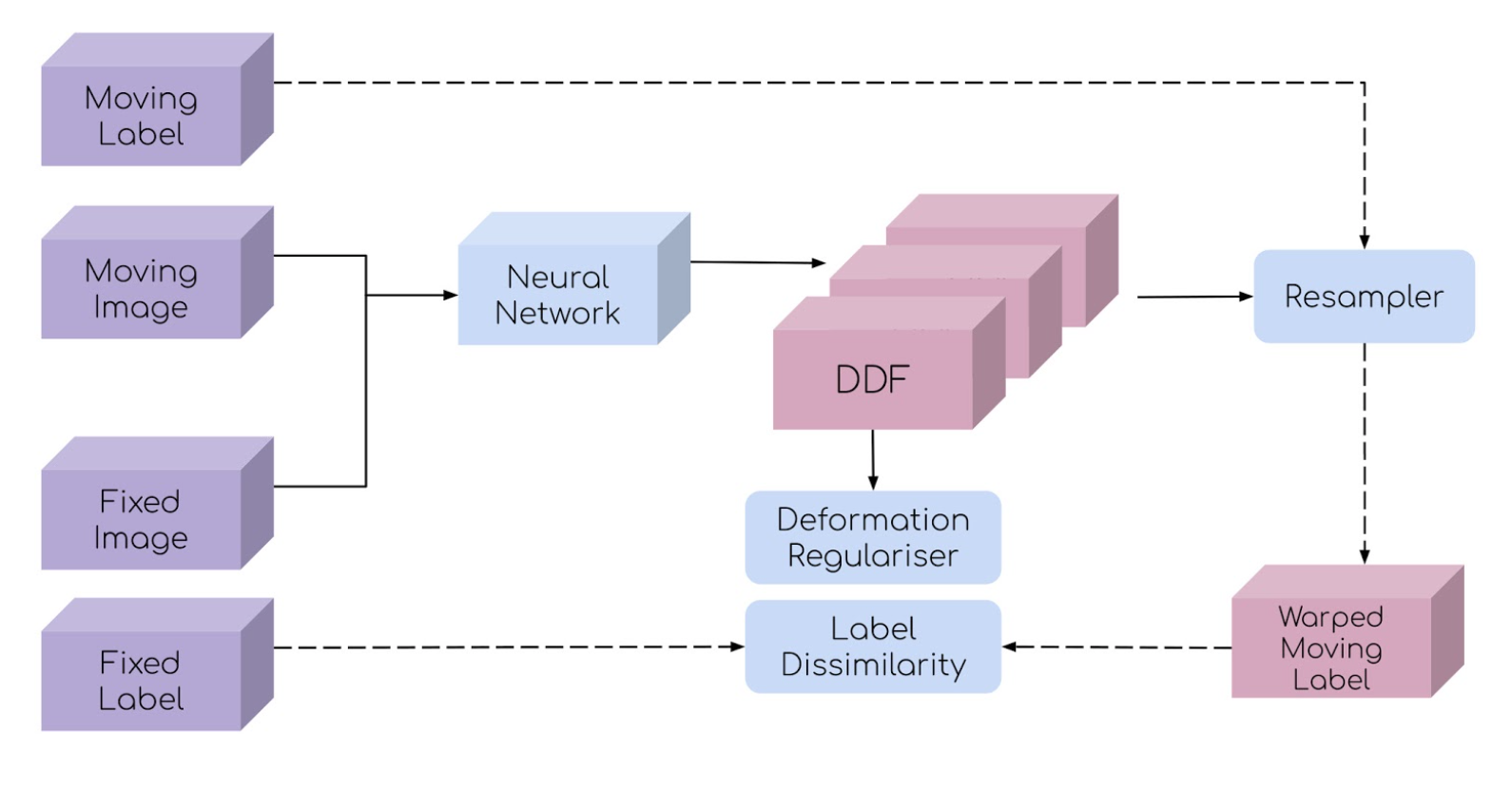}
    \caption{A schematic illustration of a weakly supervised image registration network}
    \label{fig:weak}
\end{figure}

\subsubsection{Combine}
When the data label is available, combining intensity-based, label-based, and deformation based losses together has shown superior registration accuracy, compared to unsupervised and weakly supervised methods. Following is an illustration of a combined DDF-based registration network. \par
\begin{figure}
    \centering
    \includegraphics[width=\textwidth]{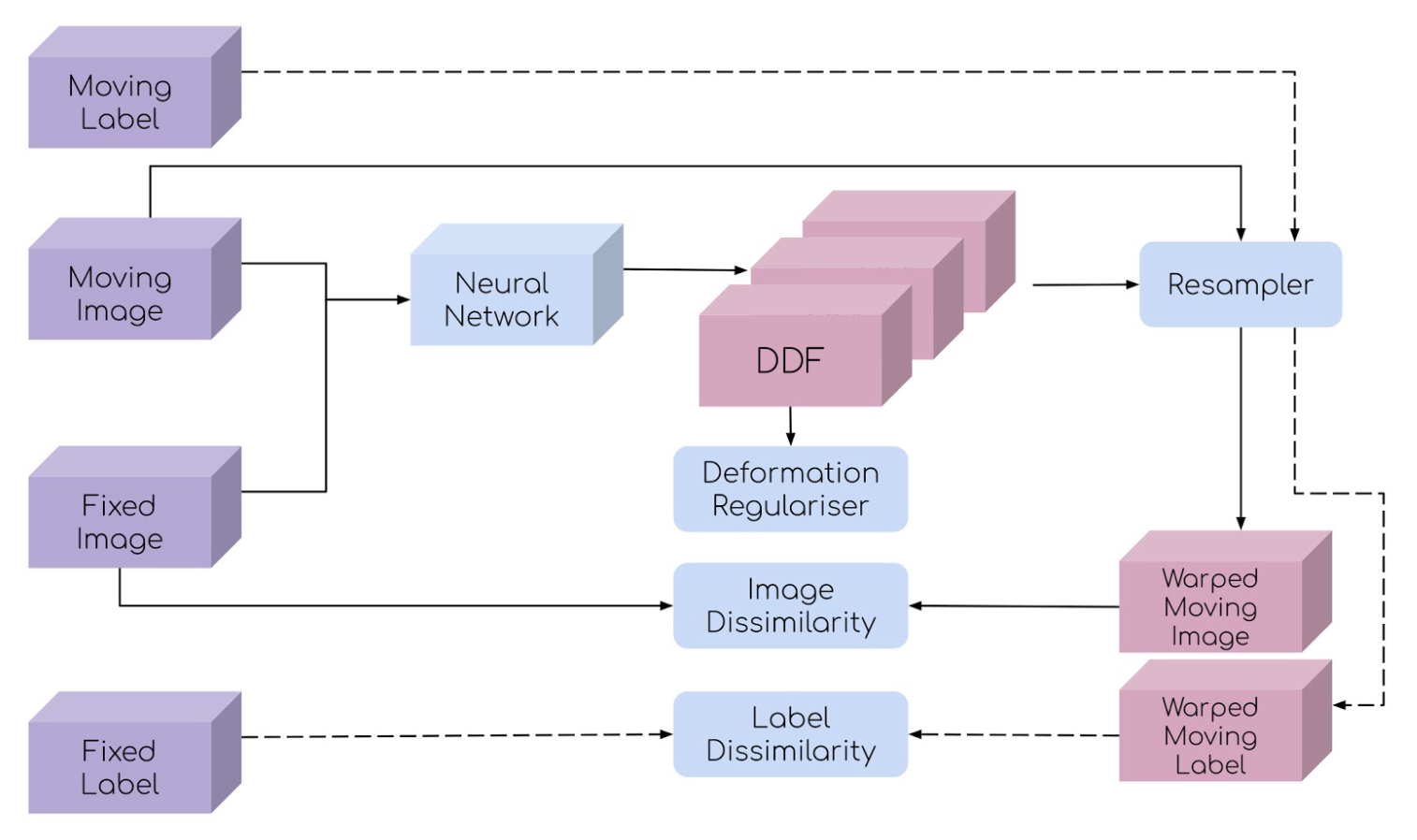}
    \caption{A schematic illustration of a combination of supervision strategies image registration network}
    \label{fig:combo}
\end{figure}

\subsubsection{A note on the relationship between feature-based registration and weakly-supervised registration}
 These segmentations, typically highlighting specific anatomical or pathological regions of interest (ROIs) in the scan(s), may also be considered a form of image features, extracted manually or using automated methods. The similarity measures or distance functions used in classical feature-based registration methods can then be used to drive the training of the weakly-supervised registration networks. These measures include the overlap between ROIs or Euclidian distance between the ROI centroids. A key insight is that the weakly-supervised learning described above is to learn the feature extraction together with the alignment of the features in an end-to-end manner.

\subsection{Loss Functions}
We aim to train a network to predict some transformation between a pair of images that is likely. To do this, we need to define what is a "likely" transformation. This is done via a loss function.\par

The loss function defined to train a registration network will depend on the type of data we have access to, yet another methodological detail drawn substantial experience from the classical methods. \par
\subsubsection{Label based loss}

Provided labels for the input images, a label based loss may be used to measure the (dis)similarity of warped regions of interest. Having computed a transformation between images using the net, one of the labels is warped and compared to the ground truth image label. Labels are typically manually contoured organs. \par
The common loss function is Dice loss, Jaccard and average cross-entropy over all voxels, which are measures of the overlap of the ROIs. For example, the Dice score between two sets, X and Y, is defined like: $$Dice=2 \frac{(X	\cap Y)}{(|X|+|Y|)}$$ \par

Let’s illustrate with some examples. We are using head and neck CT scans data \cite{PROMISE}. The data is openly accessible, the original source can be found here. \par
The labels for this data-label pair indicate the location of the spinal cord and brainstem, which typically are regions to be avoided during radiotherapy interventions. \par

\begin{lstlisting}[style = Python, numbers=none, caption={Downloading the Head-and-neck CT data}]
# Import medical data - we are going to use the head
# and neck CT data to show the losses.
MAIN_PATH = os.getcwd()
PROJECT_DIR = os.path.join(
                MAIN_PATH,
                "demos",
                "classical_ct_headandneck_affine")

DATA_PATH = "dataset"
FILE_PATH = os.path.abspath(os.path.join(PROJECT_DIR,
                                         DATA_PATH,
                                         "demo.h5"))
ORIGIN = os.path.join("https://github.com",
                      "yipenghu",
                      "example-data",
                      "raw/master/",
                      "hnct/demo.h5")

if os.path.exists(FILE_PATH):
    os.remove(FILE_PATH)   
else:
  os.makedirs(os.path.join(PROJECT_DIR, DATA_PATH))

get_file(FILE_PATH, ORIGIN)
print("CT head-and-neck data downloaded: %s."%FILE_PATH)
 
\end{lstlisting}
\begin{lstlisting}[style=Python, numbers=none, caption={Defining utility visualisation functions}]
# We define a function to visualise the results of the
# overlap for label based loss
from skimage.color  import label2rgb
def pred_label_comparison(pred,
                          mask,
                          shape_pred,
                          thresh=0.5):
    """
    Compares prediction array to mask array and returns
    rgb image with color coding corresponding to
    prediction
    outcome.
    True positive = white
    True negative = black
    False positive = green
    False negative = red

    INPUTS:
    - pred: [K M N] np.array, containing K probability
    maps
    outrue_posut from model.
    - mask: [K M N] np.array of 0s, 1s, containing K
    ground
    truths for corresponding prediction.
    - thresh: (OPT) = float between [0 1], corresponding
    to value above which to threshold predictions.

    OUTPUTS:
    - label: [K M N 3] np.array, containing K RGB images
    colour coded to show areas of intersections between
    masks and predictions.
    """
    # Create outrue_posut np.array to store images
    label = np.zeros((shape_pred[0],
                      shape_pred[1],
                      shape_pred[2],
                      3))

    # Thresholding pred
    pred_thresh = pred > thresh

    # Creating inverse to the masks and predictions
    mask_not = np.logical_not(mask)
    pred_not = np.logical_not(pred_thresh)

    # Finding intersections
    true_pos_array = np.logical_and(pred_thresh, mask)
    false_pos_array = np.logical_and(pred_thresh,
                                     mask_not)
    false_neg_array = np.logical_and(pred_not,
                                     mask)

    # Labelling via color
    false_pos_labels = 2*false_pos_array # green
    false_neg_labels = 3*false_neg_array # red
    label_array = true_pos_array + false_pos_labels
                  + false_neg_labels

    # Compare all preds to masks
    for i in range(shape_pred[0]):
        label[i, :, :, :] = label2rgb(
            label_array[i, :, :],
            bg_label=0,
            bg_color=(0, 0, 0),
            colors=[(1, 1, 1), (0, 1, 0), (1, 0, 0)])
        # 1=(tp, white), 2=(fp, green), 3=(fn, red)

    return label
 
\end{lstlisting}
\begin{lstlisting}[style=Python, numbers=none, caption={Plotting images and labels}]

# We set the plot size to some parameters.
plt.rcParams["figure.figsize"] = (10,10)

# Opening the file
fid = h5py.File(FILE_PATH, "r")

# Getting the image and label
fixed_image = tf.cast(tf.expand_dims(fid["image"],
                                     axis=0),
                      dtype=tf.float32)
fixed_labels = tf.cast(tf.expand_dims(fid["label"],
                                     axis=0),
                       dtype=tf.float32)

# Getting the 0th slice in the tensor
fixed_image_0 = fixed_image[0, ..., 0]
# Getting the 0th slice foreground label,
# at index 1 in the label tensor.
fixed_label_0 = fixed_labels[0, ..., 0, 1]
\end{lstlisting}
To compare two labels which are similar to illustrate the losses, we will slightly warp them using an affine transform. DeepReg has a set of utility functions which can be used to warp image tensors quickly. We introduce their functionalities below as we will use them throughout the rest of the tutorial.
\begin{itemize}
    \item random transform generator: generates a batch of 3D tf.Tensor affine transforms
    \item get reference grid: creates a mesh tf.Tensor of certain dimensions
    \item warp grid: using the random transform generator transforms (or other), we warp the reference mesh.
    \item resample: resamples an image/label tensor using the warped grid.
\end{itemize}

For a more in depth view of the functions refer to the documentation.\par

\begin{lstlisting}[style = Python, numbers=none, caption={Simulate warped label to illustrate label based loss}]
# Simulate a warped label
# The following function generates a random transform.
transform_random = layer_util.random_transform_generator(
                            batch_size=1,
                            scale=0.02,
                            seed=4)

# We create a reference grid of image size
grid_ref = layer_util.get_reference_grid(
                grid_size=fixed_labels.shape[1:4]
                                        )

# We warp our reference grid by our random transform
grid_random = layer_util.warp_grid(grid_ref,
                                   transform_random)
# We resample the fixed image with the random transform
# to create a distorted
# image, which we will use as our moving image.
moving_label = layer_util.resample(vol=fixed_labels,
                                   loc=grid_random
                                   )[0, ..., 0, 1]
moving_image = layer_util.resample(vol=fixed_image,
                                   loc=grid_random
                                   )[0, ..., 0]

fig, axs = plt.subplots(1, 2)
axs[0].imshow(fixed_label_0)
axs[1].imshow(fixed_image[0, ..., 0])
axs[0].set_title("Fixed label")
axs[1].set_title("Fixed image")
plt.show()
fig, axs = plt.subplots(1, 2)
axs[1].imshow(moving_image)
axs[0].imshow(moving_label>0.1)
axs[0].set_title("Moving label")
axs[1].set_title("Moving image")
plt.show()
 
 
# We set the plot size to some parameters.
plt.rcParams["figure.figsize"] = (10,10)

comparison = pred_label_comparison(
                  np.expand_dims(moving_label,
                                 axis=0),
                  np.expand_dims(fixed_label_0,
                                 axis=0),
                [1, 128, 128],
                thresh=0.1)

plt.imshow(comparison[0, :, :])
plt.title("Comparing fixed and moving label")
plt.show()

\end{lstlisting}

Where the white pixels indicate true positives, the green pixels indicate false positives (ie where the moving label has a segmentation where the fixed label does not) and the red pixels indicate false negatives (ie where the moving label lacks segmented pixels with respect to the fixed label).
Lets calculate the dice score using a function from DeepReg - it will result in a value between 0 (no overlap) and 1 (perfect overlap).

\begin{lstlisting}[style = Python, numbers=none, caption={Illustrating label based loss - calculating scores}]
from deepreg.model.loss.label import dice_score
# Calculating dice - we need [batch, dim1, dim2, dim3],
# so we expand the labels axes'
batch_moving_label = np.expand_dims(
                        np.expand_dims(moving_label,
                                       axis=0),
                        axis=-1)
batch_fixed_label = np.expand_dims(
                        np.expand_dims(fixed_label_0,
                                       axis=0),
                        axis=-1)

score_warped = dice_score(batch_moving_label,
                          batch_fixed_label)
print("Score for dissimilar labels: {}".format(
                                score_warped))
 
# We would expect the Dice Score between the same
# label to be perfect - lets check:

score_same = dice_score(batch_fixed_label,
                        batch_fixed_label)
print("Score for same labels: {}".format(score_same))
 

\end{lstlisting}

We can use this score as a measure of registration via label-driven methods such as weakly-supervised and conditional segmentation: we want to maximise the overlap score such that the two features are as similar as possible. So, to convert the score into a loss we should minimise the negative overlap measure (eg. loss = 1 - dice score) to maximise overlap of the regions during training.

\subsubsection{Intensity based (image based) loss}

This type of loss measures the dissimilarity of the fixed image and warped moving image, which is adapted from the classical image registration methods. Intensity based loss can be highly modality-dependent. The common loss functions are normalized cross correlation (NCC), sum of squared distance (SSD), and mutual information (MI) with their variants. \par
For example, the sum of square differences takes the direct difference in intensity values between moving and fixed image tensors of dimensions (batch, I, J, K, channels) as a measure of similarity by calculating the average difference per tensor: $$SSD = \frac{1}{I \times J \times K \times C}\sum\limits_{i,j,k,c}(moving_{i,j,k,c} - fixed_{i,j,k,c})^{2} $$

\begin{lstlisting}[style = Python, numbers=none, caption={Illustrating intensity based loss}]
# Illustrate intensity based loss
from deepreg.model.loss.image import ssd

# The ssd function requires
# [batch, dim1, dim2, dim3, ch] sized tensors -
# expand the image dims as ours are
# [batch, dim1, dim2, dim3]
moving_image = layer_util.resample(vol=fixed_image,
                                   loc=grid_random)

fig, axs = plt.subplots(1, 2)
axs[0].imshow(moving_image[0, ..., 0])
axs[1].imshow(fixed_image[0, ..., 0])
axs[0].set_title("Moving image")
axs[1].set_title("Fixed image")
plt.show()

# We can visualise the difference between
# tensors by calculating a new tensor
tensor_ssd = tf.square(moving_image[0, ..., 0]
             - fixed_image[0, ..., 0])
plt.imshow(tensor_ssd)
plt.show()

# We calculate over all the images
ssd_loss = ssd(np.expand_dims(moving_image,
                              axis=-1),
               np.expand_dims(fixed_image,
                              axis=-1))
print("SSD loss between the image tensors: {}".format(
                                         ssd_loss))

\end{lstlisting}

\subsubsection{Deformation loss}
Additionally, training may be regularised by computing the "likelihood" of a given displacement field. High deformation losses point to very unlikely displacement due to high gradients of the field - typically, deformation losses ensure smoothness in the displacement field. For DDFs, typical regularisation losses are bending energy losses, L1 or L2 norms of the displacement gradients.

\subsection{Image Registration With Deep Learning: Summary}
For deep learning methods, pairs of images, denoted as moving and fixed images, are passed to the network to predict a transformation between the images.
The deep learning approach for medical image registration will depend on mainly three factors:
\begin{itemize}
    \item What type of network output is one trying to predict?
    \item What type image data are being registered? Are there any other data, such as segmentations, to support the registration?
    \item Are the data paired? Are they labeled?
\end{itemize}

From this, we can design an appropriate architecture and choose an adequate loss function to motivate training.

\section{Two Classical Registration Examples}
We will use DeepReg functions to register two images.
First, we will illustrate the possibility of "self-registering" an image to it's affine-transformed counterpart, using the same head and neck CT scans data \cite{PROMISE} we used to illustrate the losses.

\subsection{Optimising an affine transformation: a "self-registration" example
}
\begin{lstlisting}[style=Python,numbers=none,  caption={Defining utility functions for optimisation}]

# We set the plot size to some parameters.
plt.rcParams["figure.figsize"] = (100,100)
 

# We define some utility functions first
## optimisation
@tf.function
def train_step_CT(grid, weights, optimizer, mov, fix):
    """
    Train step function for backprop using gradient tape.
    GradientTape is a tensorflow API which automatically
differentiates and facilitates the implementation of 
machine

    learning algorithms:
    https://www.tensorflow.org/guide/autodiff.

    :param grid: reference grid return
    from util.get_reference_grid
    :param weights: trainable affine parameters
    [1, 4, 3]
    :param optimizer: tf.optimizers: choice of optimizer
    :param mov: moving image, tensor shape
    [1, m_dim1, m_dim2, m_dim3]
    :param fix: fixed image, tensor shape
    [1, f_dim1, f_dim2, f_dim3]
    :return loss: image dissimilarity to minimise
    """
    # We initialise an instance of gradient tape
    # to track operations
    with tf.GradientTape() as tape:
        pred = layer_util.resample(
            vol=mov, 
            loc=layer_util.warp_grid(grid,
                                     weights))
        # Calculate the loss function between the
        # fixed image
        # and the moving image
        loss = image_loss.dissimilarity_fn(
            y_true=fix,
            y_pred=pred,
            name=image_loss_name
        )
    gradients = tape.gradient(loss,
                              [weights])
    # Applying the gradients
    optimizer.apply_gradients(zip(gradients, [weights]))
    return loss

def plot_results(moving_image,
                 fixed_image,
                 warped_moving_image,
                 nIdx):
  """
  Plotting the results from training
  :param moving_image: tensor dims
  [IM_SIZE_0, IM_SIZE_1, 3]
  :param fixed_image:  tensor dims
  [IM_SIZE_0, IM_SIZE_1, 3]
  :param warped_moving_image: tensor dims
  [IM_SIZE_0, IM_SIZE_1, 3]
  :param nIdx: number of indices to display
  """
  # Display
  plt.figure()
  # Generate a nIdx images in 3s
  for idx in range(nIdx):
      axs = plt.subplot(nIdx, 3, 3 * idx + 1)
      axs.imshow(moving_image[0, ..., idx_slices[idx]],
                 cmap="gray")
      axs.axis("off")
      axs = plt.subplot(nIdx, 3, 3 * idx + 2)
      axs.imshow(fixed_image[0, ..., idx_slices[idx]],
                 cmap="gray")
      axs.axis("off")
      axs = plt.subplot(nIdx, 3, 3 * idx + 3)
      axs.imshow(
        warped_moving_image[0, ..., idx_slices[idx]],
        cmap="gray")
      axs.axis("off")
  plt.ion()
  plt.suptitle(
   'Moving Image - Fixed Image - Warped Moving Image',
   fontsize=200)
  plt.show()

def display(moving_image, fixed_image):
  """
  Displaying our two image tensors to register
  :param moving_image: [IM_SIZE_0, IM_SIZE_1, 3]
  :param fixed_image:  [IM_SIZE_0, IM_SIZE_1, 3]
  """
  # Display
  idx_slices = [int(5+x*5) for x in range(
                    int(fixed_image_size[3]/5)-1)]
  nIdx = len(idx_slices)
  plt.figure()
  for idx in range(len(idx_slices)):
      axs = plt.subplot(nIdx, 2, 2*idx+1)
      axs.imshow(moving_image[0,...,idx_slices[idx]],
                 cmap='gray')
      axs.axis('off')
      axs = plt.subplot(nIdx, 2, 2*idx+2)
      axs.imshow(fixed_image[0,...,idx_slices[idx]],
                 cmap='gray')
      axs.axis('off')
  plt.suptitle('Moving Image - Fixed Image',
               fontsize=200)
  plt.show()
 
\end{lstlisting}
\begin{lstlisting}[style=Python, numbers=none, caption={Registering warped Head And Neck CT Data}]

# We re-use the data from the head and neck CT we
#used to illustrate the losses, so we don't have to
# redownload it.

## registration parameters
image_loss_name = "ssd"
learning_rate = 0.01
total_iter = int(1001)


# Opening the file
fid = h5py.File(FILE_PATH, "r")
fixed_image = tf.cast(tf.expand_dims(fid["image"],
                                     axis=0),
                      dtype=tf.float32)

# normalisation to [0,1]
fixed_image = (fixed_image - tf.reduce_min(fixed_image))
    / (
        tf.reduce_max(fixed_image)
        - tf.reduce_min(fixed_image)
      )  

# generate a radomly-affine-transformed moving image
#using DeepReg utils
fixed_image_size = fixed_image.shape
# The following function generates a random transform.
transform_random = layer_util.random_transform_generator(
                    batch_size=1, scale=0.2)

# We create a reference grid of image size
grid_ref = layer_util.get_reference_grid(
            grid_size=fixed_image_size[1:4])

# We warp our reference grid by our random transform
grid_random = layer_util.warp_grid(
                grid_ref, transform_random)
# We resample the fixed image with the
# random transform to create a distorted
# image, which we will use as our moving image.
moving_image = layer_util.resample(vol=fixed_image,
                                   loc=grid_random)

# warp the labels to get ground-truth using
# the same random affine transform
# for validation
fixed_labels = tf.cast(tf.expand_dims(fid["label"],
                                      axis=0),
                       dtype=tf.float32)
# We have multiple labels, so we apply the
# transform to all the labels by
# stacking them
moving_labels = tf.stack(
    [
        layer_util.resample(vol=fixed_labels[..., idx],
                            loc=grid_random)
        for idx in range(fixed_labels.shape[4])
    ],
    axis=4,
)

\end{lstlisting}
\begin{lstlisting}[style=Python, numbers=none, caption={Optimising the transformation}]


# We create an affine transformation as a trainable
#weight layer
var_affine = tf.Variable(
    initial_value=[
        [[1.0, 0.0, 0.0],
         [0.0, 1.0, 0.0],
         [0.0, 0.0, 1.0],
         [0.0, 0.0, 0.0]]
    ],
    trainable=True,
)

# We perform an optimisation by backpropagating the
# loss through to our 
# trainable weight layer.
optimiser = tf.optimizers.Adam(learning_rate)


# Perform an optimisation for total_iter number of steps.
for step in range(total_iter):
    loss_opt = train_step_CT(grid_ref,
                             var_affine,
                             optimiser,
                             moving_image,
                             fixed_image)
    if (step % 50) == 0:  # print info
        tf.print("Step",
                 step,
                 image_loss_name,
                 loss_opt)
\end{lstlisting}
Once the optimisation converges (this may take a minute on a GPU), we can use the optimised affine transformation to warp the moving images.
\begin{lstlisting}[style=Python, numbers=none, caption={Checking results of optimisation}]


## warp the moving image using the optimised
# affine transformation
grid_opt = layer_util.warp_grid(grid_ref, var_affine)
warped_moving_image = layer_util.resample(
                        vol=moving_image,
                        loc=grid_opt)

idx_slices = [
        int(5 + x * 5) for x in range(
                int(fixed_image_size[3] / 5) - 1)]
nIdx = len(idx_slices)
# display to check the results.
plot_results(moving_image,
             fixed_image,
             warped_moving_image,
             nIdx)
 \end{lstlisting}
We can see how the data has registered to the fixed image. Let's see how the transformation appears on the labels.
 \begin{lstlisting}[style=Python, numbers=none]
 

# Check how the labels have been registered
warped_moving_labels  = layer_util.resample(
                            vol=moving_labels,
                            loc=grid_opt)

# display
for idx_label in range(fixed_labels.shape[4]):
    plt.figure()
    for idx in range(len(idx_slices)):
        axs = plt.subplot(nIdx, 3, 3 * idx + 1)
        axs.imshow(moving_labels[0,
                                 ...,
                                 idx_slices[idx],
                                 idx_label],
                                 cmap="gray")
        axs.axis("off")
        axs = plt.subplot(nIdx, 3, 3 * idx + 2)
        axs.imshow(fixed_labels[0,
                                 ...,
                                 idx_slices[idx],
                                 idx_label],
                                 cmap="gray")
        axs.axis("off")
        axs = plt.subplot(nIdx, 3, 3 * idx + 3)
        axs.imshow(
            warped_moving_labels[0,
                                 ...,
                                 idx_slices[idx],
                                 idx_label],
                                 cmap="gray")
        axs.axis("off")
    plt.ion()
    plt.show()
  \end{lstlisting}
Here we should be able to see either or both of the following two cases. \par
\begin{itemize}
    \item There are labels appeared in some slices of the fixed and warped moving images while it does not exist in the same slices of the original moving image;
    \item Some labels in the original moving images disappeared in both fixed and warped moving images, from the same slices.
\end{itemize}

Both indicate the warped moving image has been indeed "warped" closer to the fixed image space from the moving image space.

\subsection{Optimising a nonrigid transformation: an inter-subject registration application}
Now, we will nonrigid-register inter-subject scans, using MR images from two prostate cancer patients \cite{PROMISE}. The data is from the PROMISE12 Grand Challenge. We will follow the same procedure, optimising the registration for several steps.

\begin{lstlisting}[style = Python, numbers=none,  caption={Utility Functions for optimisation}]
# Defining some utility functions
@tf.function
def train_step(warper, weights, optimizer, mov, fix):
    """
    Train step function for backpropagation using
    gradient tape.
    In contrast to CT function, we have a deformation
    regularisation.

    :param warper: warping function returned from
    layer.Warping
    :param weights: trainable ddf
    [1, f_dim1, f_dim2, f_dim3, 3]
    :param optimizer: tf.optimizers
    :param mov: moving image
    [1, m_dim1, m_dim2, m_dim3]
    :param fix: fixed image [1, f_dim1, f_dim2, f_dim3]
    :return:
        loss: overall loss to optimise
        loss_image: image dissimilarity
        loss_deform: deformation regularisation
    """
    with tf.GradientTape() as tape:
        pred = warper(inputs=[weights, mov])
        # Calculating the image loss between the
        # ground truth and prediction
        loss_image = image_loss.dissimilarity_fn(
            y_true=fix, y_pred=pred,
            name=image_loss_name
        )
        # We calculate the deformation loss
        loss_deform = (
                deform_loss.local_displacement_energy(
                            weights,
                            deform_loss_name))
                
        # Total loss is weighted
        loss = loss_image
               + weight_deform_loss * loss_deform
    # We calculate the gradients by
    # backpropagating the loss to the
    # trainable layer
    gradients = tape.gradient(loss, [weights])
    # Using our tf optimizer, we apply the gradients
    optimizer.apply_gradients(zip(gradients, [weights]))
    return loss, loss_image, loss_deform
 
\end{lstlisting}
\begin{lstlisting}[style=Python, numbers=none, caption={Downloading PROMISE data}]
## We download the data for this example.
MAIN_PATH = os.getcwd()

DATA_PATH = "dataset"
if not os.path.exists(os.path.join(MAIN_PATH,
                                   DATA_PATH)):
  os.makedirs(os.path.join(MAIN_PATH, DATA_PATH))

FILE_PATH = os.path.abspath(os.path.join(MAIN_PATH,
                                         DATA_PATH,
                                         "demo2.h5"))
ORIGIN = os.path.join("https://github.com",
                      "yipenghu",
                      "example-data/raw",
                      "master/promise12",
                      "demo2.h5")

get_file(FILE_PATH, ORIGIN)
print("Prostate MR data downloaded: %s." % FILE_PATH)

os.chdir(MAIN_PATH)

DATA_PATH = "dataset"
FILE_PATH = os.path.join(MAIN_PATH,
                         DATA_PATH,
                         "demo2.h5")

fid = h5py.File(FILE_PATH, "r")
 
\end{lstlisting}
\begin{lstlisting}[style=Python, numbers=none, caption={Defining registration parameters, and optimising transformation}]
## We define some registration parameters
# - play around with these!
image_loss_name = "lncc"
# local normalised cross correlation loss between images
deform_loss_name = "bending"
# Loss to measure the bending energy of the ddf
weight_deform_loss = 1
# we weight the deformation loss
learning_rate = 0.1
total_iter = int(3001) # This will train for longer
 

# We get our two subject images from our datasets
moving_image = tf.cast(tf.expand_dims(fid["image0"],
                                      axis=0),
                       dtype=tf.float32)
fixed_image = tf.cast(tf.expand_dims(fid["image1"],
                                     axis=0),
                      dtype=tf.float32)
 
 
# We initialise our layers
fixed_image_size = fixed_image.shape
initialiser = tf.random_normal_initializer(mean=0,
                                          stddev=1e-3)

# Creating our DDF tensor that can be trained
# The DDF will be of shape [IM_SIZE_1, IM_SIZE_2, 3],
# representing the displacement field at each pixel
# and xyz dimension.
var_ddf = tf.Variable(
            initialiser(
                fixed_image_size + [3]),
            name="ddf",
            trainable=True)

# We create a warping layer and initialise an optimizer
warping = layer.Warping(
            fixed_image_size=fixed_image_size[1:4])
optimiser = tf.optimizers.Adam(learning_rate)

## Optimising the layer
## With GPU this takes about 5 minutes.
for step in range(total_iter):
    # Call the gradient tape function
    loss_opt, loss_image_opt, loss_deform_opt =
    train_step(
      warping,
      var_ddf,
      optimiser,
      moving_image,
      fixed_image
              )           
    if (step % 50) == 0: # print at 50th step
        tf.print(
            "Step",
            step,
            "loss",
            loss_opt,
            image_loss_name,
            loss_image_opt,
            deform_loss_name,
            loss_deform_opt,
        )
 
\end{lstlisting}

\begin{lstlisting}[style=Python, numbers=none, caption={Illustrating effects of transformation}]
## Warp the moving image using the optimised ddf and
# the warping layer.
idx_slices = [
    int(5 + x * 5) for x in range(
        int(fixed_image_size[3] / 5) - 1)
              ]
nIdx = len(idx_slices)
warped_moving_image = warping(inputs=[var_ddf,
                                      moving_image])
plot_results(moving_image,
             fixed_image,
             warped_moving_image,
             nIdx)
 
 
## We can observe the effects of the warping on
# the moving label using
# the optimised affine transformation
moving_label = tf.cast(tf.expand_dims(fid["label0"],
                                      axis=0),
                      dtype=tf.float32)
fixed_label = tf.cast(tf.expand_dims(fid["label1"],
                                     axis=0),
                      dtype=tf.float32)

idx_slices = [
    int(5 + x * 5) for x in range(
        int(fixed_image_size[3] / 5) - 1)
              ]
nIdx = len(idx_slices)
warped_moving_labels = warping(inputs=[var_ddf,
                                       moving_label])
plot_results(moving_label,
             fixed_label,
             warped_moving_labels,
             nIdx)

\end{lstlisting}

\section{An Adapted DeepReg Demo}
Now, we will build a more complex demo, also a clinical case, using deep-learning.
This is a registration between CT images acquired at different time points for a single patient. The images being registered are taken at inspiration and expiration for each subject. This is an intra subject registration. This type of intra subject registration is useful when there is a need to track certain features on a medical image such as tumor location when conducting invasive procedures \cite{CTData}. \par
The data files used in this tutorial have been pre-arranged in a folder, required by the DeepReg paired dataset loader, and can be downloaded as follows.

\begin{lstlisting}[style = Python,numbers=none,  caption={Getting files for intra-subject registration}]
from tensorflow.keras.utils import get_file
import zipfile
import shutil
import os

MAIN_PATH = os.getcwd()
PROJECT_DIR = os.path.join(MAIN_PATH,
                          "demos/paired_ct_lung/")
if os.path.exists(os.path.join(PROJECT_DIR,'data')):
  shutil.rmtree(os.path.join(PROJECT_DIR,'data'))

URL_ZIP = os.path.join("https://github.com/yipenghu",
                       "example-data/archive",
                       "paired_ct_lung.zip")
                       
data_zip = get_file(os.path.abspath(
                        os.path.join(PROJECT_DIR,
                                     'data.zip')),
                        URL_ZIP)
with zipfile.ZipFile(data_zip, "r") as zf:
    zf.extractall(PROJECT_DIR)

tmp_path = os.path.join(PROJECT_DIR,
                        "example-data-paired_ct_lung")
os.rename(tmp_path, os.path.join(PROJECT_DIR,'data'))

if os.path.exists(data_zip):
    os.remove(data_zip)

print("Data downloaded and unzipped.")

\end{lstlisting}
To train a registration network is not trivial in both computational cost and, potentially, the need for network tuning. \par
The code block below downloads a pre-trained model and uses the weights to showcase the predictive power of a deep learning trained model. You can choose to pretrain your own model by running the alternative code block in the comments. The number of epochs to train for can be changed by changing $num epochs$. The default is 2 epochs but training for longer will improve performance if training from scratch.\par
Please only either run the training or the pre-trained model download. If both code blocks are run, the trained model logs will be overwritten by the pre-trained model logs.

\begin{lstlisting}[style = Python, numbers=none, caption={Training DeepReg model}]
from deepreg.train import train

######## Pre trained model ########

! git clone https://github.com/DeepRegNet/deepreg-model-zoo.git
logs/paired_ct_lung_demo_logs

import zipfile

fname = "logs" +
        "/paired_ct_lung_demo_logs/" +
        "paired_ct_lung_demo_logs.zip"

path_logs = os.path.join(
                "logs",
                "paired_ct_lung_demo_logs",
                "learn2reg_t2_paired_train_logs")

with zipfile.ZipFile(fname, "r") as zip_ref:
    zip_ref.extractall(
        r'logs/paired_ct_lung_demo_logs')

print("Files unzipped!")

if os.path.exists(path_logs) is not True:
  os.mkdir(path_logs)

! cp -rf
logs/paired_ct_lung_demo_logs/learn2reg_t2_paired_train_logs
logs

print(os.path.exists(
        r'logs/learn2reg_t2_paired_train_logs'))
 
 
######## Train from scratch, uncomment ########
# num_epochs = 2

# path_to_file = r'deepreg/config/test'

# filename = 'ddf.yaml'
# file = open(os.path.join(path_to_file,
#                          filename)
#                          ).read().splitlines()
# file[41] = file[41][:-2] + ' ' + str(num_epochs)
        
# open(os.path.join(path_to_file,
#                   filename),
#                   'w').write('\n'.join(file))

# new_file = open(os.path.join(path_to_file,
#                              filename)
#                              ).read().splitlines()
# print('Line changed to: \n', new_file[41])

# tf.test.gpu_device_name()
# print(
#    tf.config.experimental.list_physical_devices('GPU'))
# gpu = ""
# gpu_allow_growth = False
# ckpt_path = ""
# log_dir = "learn2reg_t2_paired_train_logs"
# config_path = [
#     r"deepreg/config/test/ddf.yaml",
#     r"demos/paired_ct_lung/paired_ct_lung.yaml",
# ]
# train(
#     gpu=gpu,
#     config_path=config_path,
#     gpu_allow_growth=gpu_allow_growth,
#     ckpt_path=ckpt_path,
#     log_dir=log_dir,
# )

\end{lstlisting}

With either the trained model or the downloaded model, we can predict the DDFs.
The DeepReg predict function saves images as .png files.

\begin{lstlisting}[style = Python,numbers=none]
from deepreg.predict import predict

######## Predicting Pretrained ########

log_dir = "learn2reg_t2_paired_train_logs"
ckpt_path = os.path.join("logs",
                         log_dir,
                         "save",
                         "weights-epoch100.ckpt")
config_path = os.path.join("logs",
                         "learn2reg_t2_paired_train_logs",
                         "config.yaml")

gpu = "0"
gpu_allow_growth = False
# This will take a couple of minutes
predict(
    gpu=gpu,
    gpu_allow_growth=gpu_allow_growth,
    config_path=config_path,
    ckpt_path=ckpt_path,
    mode="test",
    batch_size=1,
    log_dir=log_dir,
    sample_label="all",
)

# the numerical metrics are saved in the logs
# directory specified

######## Predicting Newly Trained ########

# log_dir = "learn2reg_t2_paired_train_logs"
# ckpt_path = os.path.join(
#                "logs",
#                log_dir,
#                "save",
#                ("weights-epoch" +
#                str(num_epochs) +
#                ".ckpt"))

# config_path = os.path.join(
#                   "logs",
#                   "learn2reg_t2_paired_train_logs",
#                   "config.yaml")

# gpu = "0"
# gpu_allow_growth = False
# predict(
#     gpu=gpu,
#     gpu_allow_growth=gpu_allow_growth,
#     config_path=config_path,
#     ckpt_path=ckpt_path,
#     mode="test",
#     batch_size=1,
#     log_dir=log_dir,
#     sample_label="all",
# )
 

\end{lstlisting}
The code block below plots different slices and their predictions generated using the trained model. The $inds to plot$ variable can be changed to plot more slices or different slices.

\begin{lstlisting}[style = Python, numbers=none]
import matplotlib.pyplot as plt
%matplotlib inline
plt.rcParams['figure.figsize'] = [20, 20]


######## Visualisation ########

# Now lets load in a few samples from the predictions
# and plot them

# change the following line to the path to image0 label0
path_to_image0_label0 = (
        r"logs/learn2reg_t2_paired_train_logs/test"
                        )

path_to_fixed_label = os.path.join(
                        path_to_image0_label0,
                        r"pair_1_label_0/fixed_label")
path_to_fixed_image = os.path.join(
                        path_to_image0_label0,
                        r"pair_1_label_0/fixed_image")

path_to_moving_label = os.path.join(
                        path_to_image0_label0,
                        r"pair_1_label_0/moving_label")
path_to_moving_image = os.path.join(
                        path_to_image0_label0,
                        r"pair_1_label_0/moving_image")


path_to_pred_fixed_image = os.path.join(
                            path_to_image0_label0,
                            r"pair_1_label_0/pred_fixed_image")
path_to_pred_fixed_label = os.path.join(
                        path_to_image0_label0,
                        r"pair_1_label_0/pred_fixed_label")

# change inds_to_plot if different images need to be
# plotted instead

inds_to_plot = [144, 145, 184, 140, 150, 180]
sub_plot_counter = 1

for ind in inds_to_plot:
  plt.subplot(6, 8, sub_plot_counter)
  label = plt.imread(
            os.path.join(path_to_fixed_label, 
                        "depth" +
                        str(ind) +
                        "_fixed_label.png"))
  plt.imshow(label)
  plt.axis("off")
  if sub_plot_counter == 1:
    plt.title("Fixed Label")

  plt.subplot(6, 8, sub_plot_counter + 1)
  fixed_im = plt.imread(
                os.path.join(path_to_fixed_image,
                            "depth" +
                            str(ind) +
                            "_fixed_image.png"))
  plt.imshow(fixed_im)
  plt.axis("off")
  if sub_plot_counter == 1:
    plt.title("Fixed Image")

  plt.subplot(6, 8, sub_plot_counter + 2)
  moving_label = plt.imread(
                    os.path.join(path_to_moving_label, 
                                "depth" +
                                str(ind) +
                                "_moving_label.png"))
  plt.imshow(moving_label)
  plt.axis("off")
  if sub_plot_counter == 1:
    plt.title("Moving Label")

  plt.subplot(6, 8, sub_plot_counter + 3)
  moving_im = plt.imread(
                os.path.join(path_to_moving_image,
                            "depth" +
                            str(ind) +
                            "_moving_image.png"))
  plt.imshow(moving_im)
  plt.axis("off")
  if sub_plot_counter == 1:
    plt.title("Moving Image")


  plt.subplot(6, 8, sub_plot_counter + 4)
  pred = plt.imread(
            os.path.join(path_to_pred_fixed_label,
                        "depth" +
                        str(ind) +
                        "_pred_fixed_label.png"))
  plt.imshow(pred)
  plt.axis("off")
  if sub_plot_counter == 1:
    plt.title("Warped Moving Label")

  plt.subplot(6, 8, sub_plot_counter + 5)
  pred_fixed_im = plt.imread(
                    os.path.join(path_to_pred_fixed_image,
                                "depth" +
                                str(ind) +
                                "_pred_fixed_image.png"))
  plt.imshow(pred_fixed_im)
  plt.axis("off")
  if sub_plot_counter == 1:
    plt.title("Warped Moving Image")

  plt.subplot(6, 8, sub_plot_counter + 6)
  pred_label = pred_label_comparison(np.expand_dims(
                                        pred[:, :, 0],
                                        axis=0),
  np.expand_dims(label[:, :, 0], axis=0), 
  (1, pred.shape[0], pred.shape[1], pred.shape[2]))
  plt.imshow(np.squeeze(pred_label))
  plt.axis("off")
  if sub_plot_counter == 1:
    plt.title("Comparison")

  plt.subplot(6, 8, sub_plot_counter + 7)
  plt.imshow(np.squeeze((-fixed_im[:, :, 0:3] +
                         pred_fixed_im[:,:, 0:3])))
  plt.axis("off")
  if sub_plot_counter == 1:
    plt.title("Pred - Fixed")   
  sub_plot_counter = sub_plot_counter + 8

plt.show()
 

\end{lstlisting}

\section{Concluding Remarks}
In this tutorial, we use two classical image registration algorithms and a deep-learning registration network, all implemented with DeepReg, to discuss the basics of modern image registration algorithms. In particular, we show that they share principles, methodologies and code between the old and the new.\par
DeepReg is a new open-source project that has a unique set of principles aiming to consolidate the research field of medical image registration, open, community-supported and clinical-application-driven. It is these features that have motivated efforts such as this tutorial to communicate with wider groups of researchers and to facilitate diverse clinical applications. \par
This tutorial may serve as a starting point for the next generation of researchers to have a balanced starting point between the new learning-based methods and classical algorithms. It may also be used as a quick introduction of DeepReg to those, who have significant experience in deep learning and / or medical image registration, such that they can make an informed judgement whether this new tool can help their research.

%
% ---- Bibliography ----
%
% BibTeX users should specify bibliography style 'splncs04'.
% References will then be sorted and formatted in the correct style.
%
\bibliographystyle{splncs04}
% \bibliography{mybibliography}

\begin{thebibliography}{8}

\bibitem{Haskins20}
G. Haskins, U. Kruger, and P. Yan, “Deep learning in medical image registration: a survey,” Mach. Vis. Appl., vol. 31, no. 1, pp. 1–18, Jan. 2020.
\bibitem{Hu12}
Y. Hu et al., “MR to ultrasound registration for image-guided prostate interventions,” Med. Image Anal., vol. 16, no. 3, pp. 687–703, Apr. 2012.

\bibitem{Ramalhinho18}
J. Ramalhinho et al., “A pre-operative planning framework for global registration of laparoscopic ultrasound to CT images,” Int. J. Comput. Assist. Radiol. Surg., vol. 13, no. 8, pp. 1177–1186, Aug. 2018.

\bibitem{Valdes02}
M. Lorenzo-Valdés, G. I. Sanchez-Ortiz, R. Mohiaddin, and D. Rueckert, “Atlas-based segmentation and tracking of 3D cardiac MR images using non-rigid registration,” in Lecture Notes in Computer Science, 2002, vol. 2488, pp. 642–650.

\bibitem{Cazoulat16}
G. Cazoulat, D. Owen, M. M. Matuszak, J. M. Balter, and K. K. Brock, “Biomechanical deformable image registration of longitudinal lung CT images using vessel information,” Phys. Med. Biol., vol. 61, no. 13, pp. 4826–4839, 2016.

\bibitem{Hu15}
Y. Hu, et al., “Population-based prediction of subject-specific prostate deformation for MR-to-ultrasound image registration,” Med. Image Anal., vol. 26, no. 1, pp. 332–344, Dec. 2015.

\bibitem{Besl92}
P. J. Besl and N. D. McKay, “Method for registration of 3-D shapes,” in Sensor Fusion IV: Control Paradigms and Data Structures, 1992, vol. 1611, pp. 586–606.

\bibitem{Myronenko10}
A. Myronenko and X. Song, “Point set registration: Coherent point drifts,” IEEE Trans. Pattern Anal. Mach. Intell., vol. 32, no. 12, pp. 2262–2275, 2010.

\bibitem{Ashburner07}
J. Ashburner, “A fast diffeomorphic image registration algorithm,” Neuroimage, vol. 38, no. 1, pp. 95–113, Oct. 2007.

\bibitem{Hu19}
Y. Hu, E. Gibson, D. C. Barratt, M. Emberton, J. A. Noble, and T. Vercauteren, “Conditional Segmentation in Lieu of Image Registration,” in Lecture Notes in Computer Science (including subseries Lecture Notes in Artificial Intelligence and Lecture Notes in Bioinformatics), 2019, vol. 11765 LNCS, pp. 401–409.

\bibitem{Hill01}
D. L. G. Hill, P. G. Batchelor, M. Holden, and D. J. Hawkes, “Medical image registration,” Phys. Med. Biol., vol. 46, no. 3, pp. R1–R45, Mar. 2001.

\bibitem{Vallieres17}
Vallières, M. et al. Radiomics strategies for risk assessment of tumour failure in head-and-neck cancer. Sci Rep 7, 10117 (2017). doi: 10.1038/s41598-017-10371-5

\bibitem{PROMISE}
Litjens, et al., 2014. Evaluation of prostate segmentation algorithms for MRI: the PROMISE12 challenge. Medical image analysis, 18(2), pp.359-373.

\bibitem{CTData}
A. Hering, K. Murphy, and B. van Ginneken. Lean2Reg Challenge: CT Lung Registration - Training Data [Data set]. Zenodo. http://doi.org/10.5281/zenodo.3835682. 2020

\bibitem{Vercauteren09}
T. Vercauteren, et al. Diffeomorphic demons: Efficient non-parametric image registration. NeuroImage, 45(1), pp.S61-S72. 2009.
\bibitem{Yang20}
Q. Yang, et al., Longitudinal image registration with temporal-order and subject-specificity discrimination. MICCAI 2020, 2020.

\end{thebibliography}
%

\end{document}